# Elastic Waves Scattering without Conversion in Metamaterials with Simultaneous Zero Indices for Longitudinal and Transverse waves

Fengming Liu,[1,3*] and Zhengyou Liu[2§]

[1]School of Science, Hubei University of Technology, Wuhan 430068, China

[2]School of Physics and Technology, and Institute for Advanced Studies, Wuhan University, Wuhan 430072, China

[3]Hubei Collaborative Innovation Center for High-efficiency Utilization of Solar Energy, Hubei University of Technology, Wuhan, 430068, China

[*]Corresponding author:   fmliu@mail.hbut.edu.cn

[§]Corresponding author:   zyliu@whu.edu.cn

PACS numbers: 81.05.Xj, 46.40.Cd, 62.30.+d

**Abstract**

We theoretically investigate elastic waves propagating in metamaterials with simultaneous zero indices for both the longitudinal and transverse waves. With scattering objects (here cylinders) present in the metamaterials slabs, while the elastic waves can mostly transmit through the metamaterials slabs perfectly, exhibiting the well-known cloaking effect of zero index metamaterials, they nevertheless become totally blocked at resonances, indicating strong elastic waves scattering by the objects in the cases. However, despite the occurrence of the elastic waves scattering, there is, counter-intuitively, no mode conversion between the longitudinal and transverse waves in the process, completely in contrast with that in conventional elastic media. A design of two-dimensional phononic crystal with these peculiar properties is presented.

In the last decade, zero-index-metamaterials (ZIM) have become an attractive research focus. This is the category of metamaterials whose permittivity and permeability are simultaneously or individually near zero. As a consequence of the zero refractive index, the phase velocity of wave in ZIM can approach infinity, thus the phase of wave throughout the whole ZIM is essentially constant. This unique property leads to many intriguing phenomena and applications, such as tailoring the phase pattern of radiation field [1-3], tunneling of EM energy through ultra thin channels or bends [4,5], and manipulating EM wave propagation through ZIM waveguides by tailoring the parameters of the dielectric defects [6,7]. Meanwhile, the concept of metamaterials has been extended to acoustic and elastic media. Much effort has been focused on negative index of refraction [8-12], sub-wavelength imaging [13, 14], and transformation acoustic [15-21]. Recently, acoustic ZIM have also drawn intense attention and various schemes have been proposed to realize them, such as acoustic waveguides loaded with membranes and/or Helmholtz resonator [22-24], coiling up space with curled channels [25, 26], and two-dimensional (2D) acoustic crystal with Dirac-like cones dispersion [27-30]. However, because of the complexity of the scattering of elastic waves in solid structure, limited works were devoted to elastic ZIM [31, 32].

As is well known, when an elastic wave of either longitudinal (P) or transverse (S) type is incident on an elastic discontinuity, it undergoes scattering, and the scattered waves of both types are generally produced, a process known as "mode conversion" [33]. Therefore, in dealing with the 2D scattering problems of in-plane elastic waves, one needs to consider the coupled P and S waves even if in the elastic metamaterials [15, 16]. However, here we show that, in an elastic metamaterials possessing near zero reciprocal of shear modulus $1/\mu$, near zero mass density $\rho$ and ordinary bulk modulus $\kappa$ at certain frequency, wave scattering vanishes mostly for embedded objects, and even if scattering occurs at resonances, mode conversion does not happen, so that the wave natures of both the P waves and S waves always remain intact. Note that, when $\mu$ diverges, although $\kappa$ has finite value, $(\mu+\kappa)$ also diverges and $1/(\mu+\kappa)$ also tends to zero for P wave. Consequently, this kind of metamaterials has simultaneous zero indices for both the P wave and S wave, which is termed double zero-index-metamaterials (DZIM) hereafter. In the meantime, because the DZIM having its material parameters $\rho$, $1/\mu$ and $1/(\mu+\kappa)$ simultaneously go through zero at the same frequency point, and thus all of them can be expanded in a Taylor series of $\Delta\omega$ (reference to this zero point), and the first order term of $\Delta\omega$ will take the lead since generally the parameters of a metamaterial can be described with a (generalized) Drude model and the first order terms in expansions do not vanish. Taking the leading term only in the expansions in approximation, substituting them into the dispersion relations $k=\omega\sqrt{\rho/\mu}$ (for the S waves) and $k=\omega\sqrt{\rho/(\mu+\kappa)}$ (for the P wave), and considering $\omega$ constant around the zero point, will immediately lead to two linear dispersion relations for the S and P waves, which form a double Dirac cone with the Dirac point just locating at the zero point. Therefore in addition to the double zero indices, generally the DZIMs have also a double Dirac cone. As the analog of the ZIMs for electromagnetic waves and for acoustic waves, naturally one expects that the DZIM can serve as cloak for elastic waves, if its impedance matches with the surrounding medium. However, it is interesting to note that for a DZIM slab with embedded objects (e.g., cylinders), strong scattering occurs at resonances, resulting in a total reflection of the incident wave by the slab. Although scattering does occur, there is no mode conversion in the process. A simple analytic model is proposed to capture the physics of this anomalous phenomenon. In addition, we propose a 2D phononic crystal (PC) which can be mapped to an isotropic elastic material with effective zero $1/\mu_{eff}$ and zero $\rho_{eff}$. Numerical simulations show that the PC system can be a good candidate to achieve the DZIM structure experimentally for its simple manufacturing requirements and no demand of any anisotropic material parameters.

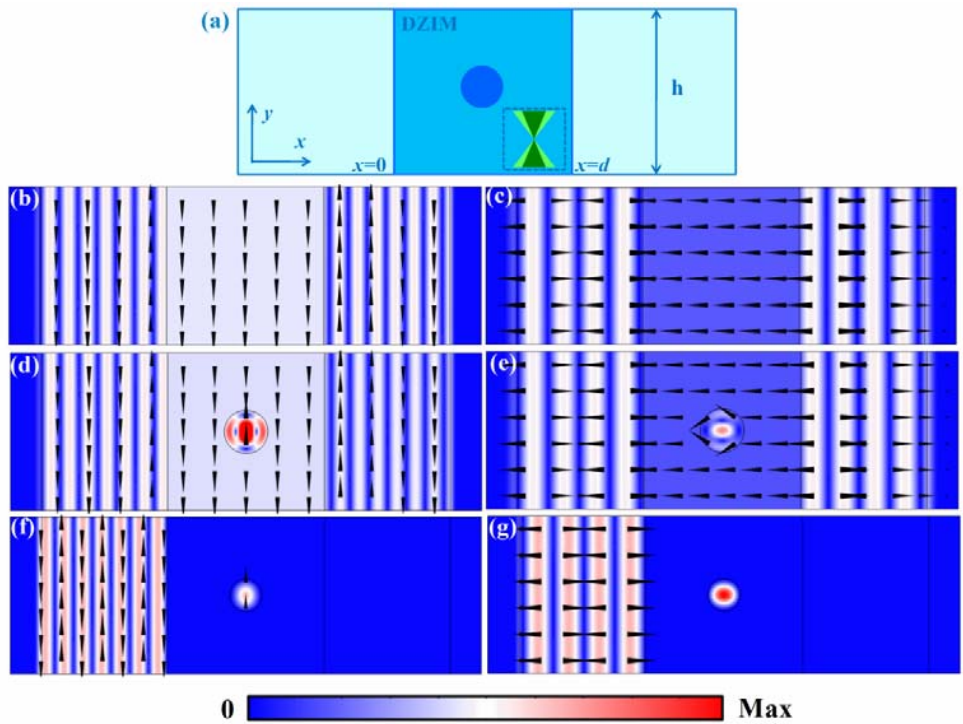

FIG. 1. (a) Schematic of the unit cell of our periodic system along y direction; the unit cell consists of the background medium, the elastic DZIM, and an embedded cylinder. The inset shows schematically the double Dirac cones dispersion of the elastic DZIM

with the slopes exaggerated for visibility. The numerically simulated displacement field distributions of the incident waves transmitting through the DZIM slab without inclusion (panels (b) and (c)), with steel inclusions (panels (d) and (e)), and the incident waves totally blocked (panels (f) and (g)). The arrows denote the directions of the displacement. Different columns represent the incident S and P waves, respectively. The simulation domain is terminated in the propagation direction with perfectly matched layers (PML) and periodic boundary condition is assumed on the upper and lower boundary.

The geometry of the 2D structure under consideration is illustrated in Fig. 1(a). It consists of four distinct regions: The left and right regions are background medium (with mass density $\rho_0$, bulk modulus $\kappa_0$, and shear modulus $\mu_0$) and are separated by the elastic DZIM slab with effective mass density $\rho_1$, bulk modulus $\kappa_1$, and shear modulus $\mu_1$. And a cylindrical solid object with radius $r_d$, mass density $\rho_d$, bulk modulus $\kappa_d$, and shear modulus $\mu_d$ is embedded in the DZIM slab. The periodic boundary condition is applied to the upper and lower boundaries in the simulation. Because of possessing double near zero parameters for both P wave and S wave, the DZIM displays double Dirac cones dispersion at certain finite frequency.

We first investigate the propagation characteristics of the DZIM slab without inclusion. The basic field equations for the in-plane elastic waves in an isotropic solid can be written as [33]

$$\rho \frac{\partial v_x}{\partial t} = \frac{\partial \tau_{xx}}{\partial x} + \frac{\partial \tau_{yx}}{\partial y}$$

$$\rho \frac{\partial v_y}{\partial t} = \frac{\partial \tau_{xy}}{\partial x} + \frac{\partial \tau_{yy}}{\partial y} \quad (1)$$

$$\frac{\partial \tau_{xx}}{\partial t} = (\kappa + \mu)\frac{\partial v_x}{\partial x} + (\kappa - \mu)\frac{\partial v_y}{\partial y}$$

$$\frac{\partial \tau_{yy}}{\partial t} = (\kappa - \mu)\frac{\partial v_x}{\partial x} + (\kappa + \mu)\frac{\partial v_y}{\partial y}$$

$$\frac{\partial \tau_{xy}}{\partial t} = \mu(\frac{\partial v_x}{\partial x} + \frac{\partial v_y}{\partial y}) \quad (2)$$

where $v_{i(i=x,y)}$ represents the velocity field which is the derivative of the displacement field $u_{i(i=x,y)}$ with respect to time, and $\tau_{ij(i,j=x,y)}$ represents the stress tensor. Eq. (1) and Eq. (2) represent Newton's law and generalized Hook' law, respectively. For plane waves propagating in homogeneous elastic medium along *x* direction, Eq. (1) and Eq. (2) can be simplified to

$$\rho \frac{\partial v_y}{\partial t} = \frac{\partial \tau_{xy}}{\partial x}, \frac{\partial \tau_{xy}}{\partial t} = \mu \frac{\partial v_y}{\partial x}, \quad (3)$$

for S wave and

$$\rho \frac{\partial v_x}{\partial t} = \frac{\partial \tau_{xx}}{\partial x}, \frac{\partial \tau_{xx}}{\partial t} = (\kappa + \mu)\frac{\partial v_x}{\partial x} \quad (4)$$

for P wave, respectively. In the DZIM region, as $1/\mu_1$ tends to zero, the velocity field $v_y$ in Eq. (3) must be constant to keep $\tau_{xy}$ as finite value. Since the displacement field is the integral of the velocity field with respect to time, $u_y$ is also constant in the DZIM region for S wave incidence. While for P wave incidence, as $1/(\mu_1 + \kappa_1)$ tends to zero too, the velocity field $v_x$ (and thus displacement field $u_x$) in Eq. (4) must be constant to keep $\tau_{xx}$ as finite value. Numerical simulations are carried out by using the finite element method (FEM) to verify the analysis. In the simulation, the background medium is Si with $\rho_0 = 2.53 \times 10^3 \, kg/m^3$, $c_{0,l} = 6.72 \times 10^3 \, m/s$ and $c_{0,t} = 4.123 \times 10^3 \, m/s$ and we have set $\rho_1 = 0.0001\rho_0$, $1/\mu_1 = 0.0001(1/\mu_0)$, and $\kappa_1 = \kappa_0$. The frequency of the incident wave is $f_0 = 863\text{Hz}$. Figs. 1(b) and 1(c) show the displacement field distributions of S wave and P wave transmitting through the DZIM slab, respectively. The displacement fields $u_{1y}$ and $u_{1x}$ in the DZIM region are uniform indeed. Next, we consider the problem of introducing solid inclusion into the DZIM slab. Figs. 1(d) and 1(e) show the numerically simulated displacement field distributions of S wave and P wave transmitting through the DZIM slab embedded with steel cylinders ($r_d = 2.8\text{m}$), respectively. Compared to Figs. 1(b) and 1(c), the displacement field distributions outside the cylinders are the same. So, even though inclusions have been introduced, the displacement fields $u_{1y}$ and $u_{1x}$ in the DZIM region are still uniform and there is no scattering and no mode conversion occurring for both types of incident waves. This phenomenon can be explained by utilizing the constitutive relation of isotropic elastic solid under which condition $\mu_1$ diverges in the DZIM region. In general, if there are elastic discontinuities one needs to consider the complete equation of generalized Hook' law (Eq. (2)) instead of Eq. (3) and Eq. (4). Since both $v_y$ and $v_x$ appear in Eq. (2), P wave and S wave are coupled. However, in the DZIM region, after both sides are divided by $\mu_1$ and under the condition $\mu_1$ diverge, Eq. (2) can be simplified to

$$0 = \frac{\partial v_{1x}}{\partial x} - \frac{\partial v_{1y}}{\partial y}$$
$$0 = -\frac{\partial v_{1x}}{\partial x} + \frac{\partial v_{1y}}{\partial y}$$
$$0 = \frac{\partial v_{1x}}{\partial x} + \frac{\partial v_{1y}}{\partial y} \quad . \tag{5}$$

Both solutions $v_{1y}$=constant, $v_{1x}$=0 (for purely S wave) and $v_{1x}$=constant, $v_{1y}$=0 (for purely P wave) can satisfy Eq. (5). Thus, purely S wave or purely P wave can still propagate in the DZIM region even though elastic discontinuities have been introduced. However, when the embedded cylinders' size is increased to have radius $r_d$=2.06m, contrary to our expectation, the incident waves are totally blocked by the DZIM slab, as shown in Figs. 1(f) and 1(g). Such anomalous total blocking is completely counterintuitive, considering the wavelengths in the DZIM approach infinity and its impedance matches with that of the background medium. What followed, a simple analytic model [34] is proposed to capture the essence of the physics. Let us consider using the case of incident P wave as an example, the derivation for incident S wave is similar. Suppose a plane harmonic P wave $u_x^{inc} = u_x e^{i(k_0 x - \omega t)}$ is incident from left into the unit cell presented in Fig. 1(a), where $u_x$ is the amplitude of the incident field, $k_0$ is the wave vector in background medium, and $\omega$ is the angular frequency. We omit the time variation item in the rest of this paper for convenience. Thus, the displacement field in the left background region can be written as

$$u_{0x} = u_x [e^{ik_0 x} + \text{R} e^{-ik_0 x}], \quad (6)$$

while in the right background region the displacement field must have the form

$$u_{1x} = u_x T e^{ik_0 (x-d)}, \quad (7)$$

where $R$ and $T$ are the reflection and transmission coefficients. In the DZIM region, the displacement field maintains a quasi-static situation ($u_{1x}$=constant regardless whether there are objects). Then using the continuous boundary condition at $x=d$, we have $u_x T = u_{1x}$, thus $T = u_{1x}/u_x$. Obviously, $T$=0 (total blocking) occurs if $u_{1x}$=0 which means that the displacement field disappears anywhere in the DZIM region. In fact, it can be seen from Fig. 1(g) that the displacement field $u_{1x}$ inside the DZIM region disappears indeed. And then a natural question to be asked is how to obtain $u_{1x}=0$. In the solid cylinders, the displacement field $u_d$ is described by the elastic wave equation

$$(\lambda_d + 2\mu_d)\nabla(\nabla \cdot u_d) - \mu_d \nabla \times \nabla \times u_d + \rho_d \omega^2 u_d = 0, \qquad (8)$$

where $\lambda_d$ represents the Lame constant satisfying $\kappa_d = \lambda_d + \mu_d$. The mirror symmetry about the $x$ axis indicates that the displacement field in the solid cylinder may be expressed in terms of potential functions as

$$u_d = \nabla \Phi + \nabla \times (\hat{z}\Psi). \qquad (9)$$

The solutions for $\Phi$ and $\Psi$ may be written as

$$\Phi = \sum_{n=0}^{\infty} A_{n1} J_n(k_{dl} r) \cos(n\theta), \qquad (10)$$

$$\Psi = \sum_{n=0}^{\infty} A_{n2} J_n(k_{dt} r) \sin(n\theta), \qquad (11)$$

where $k_{dl} = \omega\sqrt{\rho_d/(\kappa_d + \mu_d)}$, $k_{dt} = \omega\sqrt{\rho_d/\mu_d}$, $J_n(x)$ is the $n$th order Bessel function. The displacement continuity at the cylinder boundary requires that in the radial direction $u_{dr}|_{r=r_d} = u_{1x}\cos\theta$, $\qquad (12)$

and in the tangential direction

$$u_{d\theta}|_{r=r_d} = -u_{1x}\sin\theta, \qquad (13)$$

which means that in Eq. (10) and Eq. (11), we have to keep only the terms with $n=1$ to produce the necessary $\theta$ dependence. This leads to a set of linear equations

$$\begin{aligned} E_{11}A_{11} + E_{12}A_{12} &= r_d u_{1x} \\ E_{11}A_{11} + E_{12}A_{12} &= r_d u_{1x}, \end{aligned} \qquad (14)$$

where $E_{ij}$ are defined as $E_{11} = k_{dl} r_d J_0(k_{dl} r_d) - J_1(k_{dl} r_d)$, $E_{12} = J_1(k_{dt} r_d)$, $E_{21} = J_1(k_{dl} r_d)$, $E_{22} = k_{dt} r_d J_0(k_{dt} r_d) - J_1(k_{dt} r_d)$. Solving Eq. (14), we obtain

$$u_{1x} = \frac{E_{11}E_{22} - E_{12}E_{21}}{(E_{22} - E_{12})r_d} A_{11} = \frac{E_{11}E_{22} - E_{12}E_{21}}{(E_{11} - E_{21})r_d} A_{12}. \tag{15}$$

From Eq. (15), we can see that if $E_{11}E_{22} - E_{12}E_{21} = 0$, then $u_{1x}=0$ and thus *T*=0, in which case the total blocking happens. For incident S wave, after a similar processing, we find the same condition ($E_{11}E_{22} - E_{12}E_{21} = 0$) needs to be satisfied to obtain *T*=0. It is worth noting that the condition is unrelated to the structure period *h*, which is reasonable for the displacement field in the DZIM region disappears when the total blocking occurs and thus the multiple scattering among the cylinders does not need to be considered. As an example, steel cylinders are considered here. Figs. 2(a) and 2(b) show the values of expression $E_{11}E_{22} - E_{12}E_{21}$ and the numerically calculated transmission coefficients as a function of the radius $r_d$ of the cylinder, respectively. It can be seen that each time the expression $E_{11}E_{22} - E_{12}E_{21}$ equals zero, the transmission coefficients of both incident S and P waves equal zero indeed. It should be noted that here both the P and S waves are controlled by the same factor because of the high symmetry of the cylindrical object (the resonances induced by the vertical vibration of the P wave and by the horizontal vibration of the S wave are equivalent). If we use object with lower symmetry, to replace the cylindrical object here, the P and S waves can be controlled by different factors, and independent control of P waves and S waves can be realized. The independent control of P and S waves by rectangular objects, which shows good performance, is presented in the supplementary material [35]. It can also be seen in Fig. 2(b) that even we double the structure period *h*, total blocking still occurs at the same radii of the cylinder. As a result, we may expect that even single cylinder can be used to achieve the total blocking of a large DZIM region.

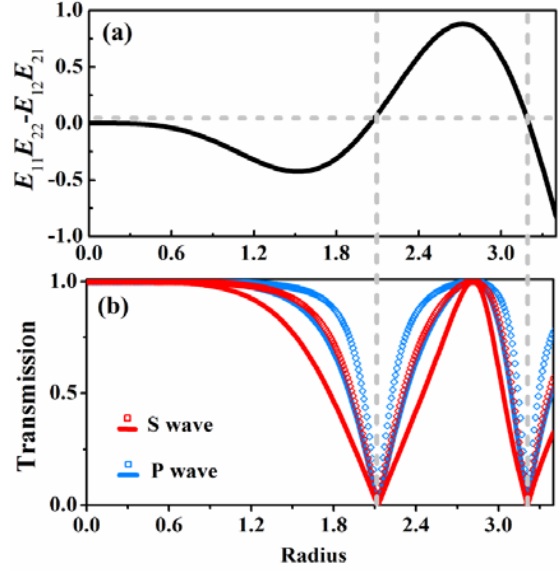

FIG. 2. The values of expression $E_{11}E_{22} - E_{12}E_{21}$ (a) and the numerically calculated transmission coefficients (b) as a function of the radius $r_d$ of the cylinder, respectively. In panel (b), the solid and square lines represent the systems with different structure periods $h$=30m and $h$=60m, respectively.

Fig. 3, as an example, shows a switchable device that can block the incident waves or tailor the radiation phase pattern. The device has rectangular geometry and is made of the DZIM and a single embedded inclusion. In Figs. 3 (a) and 3(b), a steel cylinder with radius $r_d$=2.06m (satisfying the total blocking condition) is embedded into the DZIM rectangle, either S wave or P wave incident Gaussian beam from the bottom is totally reflected. While Figs. 3(c) and 3(d) show that if the radius of the cylinder is changed to $r_d$=2.8m dissatisfying the total blocking condition, the incident wave may be transformed into radiation waves with desired shape. For S wave incidence (Fig. 3(c)), the rectangle as a whole vibrates in the horizontal direction since the displacement field is uniform in the DZIM region, and therefore the radiation wave is plane P wave in the horizontal direction while it is plane S wave in the vertical direction.

While for P wave incidence (Fig. 3(d)), the rectangle as a whole vibrates in the vertical direction, and therefore the types of radiation waves exchange in the two directions.

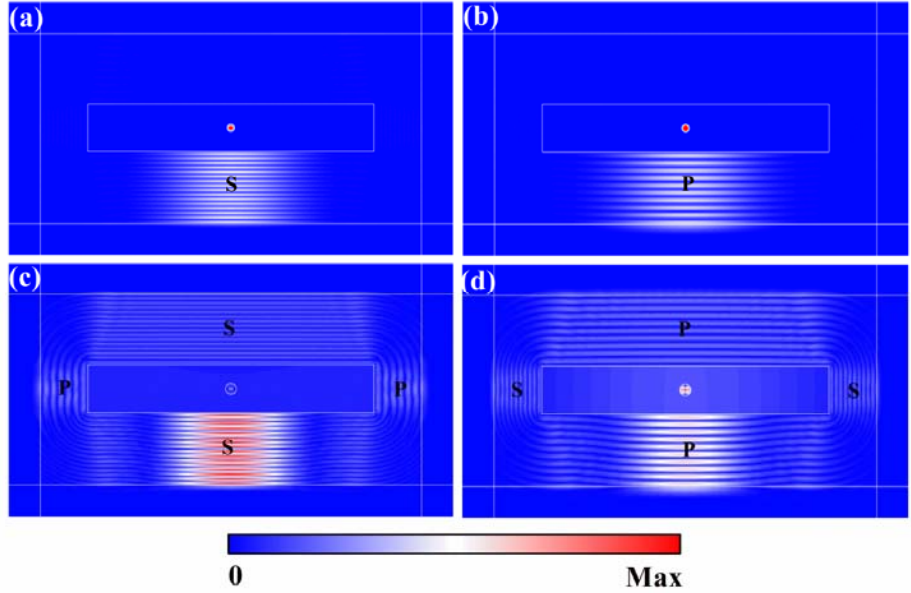

FIG. 3. Different columns represent the incident S and P waves, respectively. The numerically simulated displacement field distributions for the switchable device realizing total blocking (panels (a) and (b)) and radiation phase pattern tailoring (panels (c) and (d)). The radii of the cylinders are $r_d = 2.06$m and $r_d = 2.8$m, respectively. All sides of the simulation domain are surrounded by PML.

Finally, we investigate the experimental feasibility of the theoretical proposal. A 2D PC is designed to have double Dirac cones dispersion at the zone center and the details of the PC are presented in the supplementary material [35].

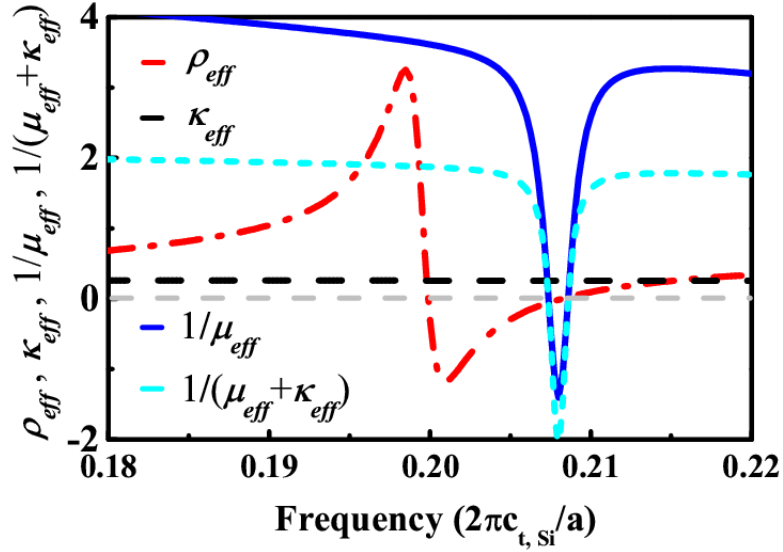

FIG. 4. The effective mass density $\rho_{eff}$, bulk modulus $\kappa_{eff}$, reciprocal of shear modulus $1/\mu_{eff}$, and $1/(\mu_{eff}+\kappa_{eff})$ as a function of frequency near the double Dirac point frequency. All the effective elastic parameters have been normalized to the elastic parameters of Si. The quantity $c_{t,Si}$ is the velocity of transverse waves of Si. The lattice constant is denoted by *a*.

We note that the energy associated with displacement fields of the eigenstates mainly localize in the rubber cylinders (the wave velocity of rubber is lower than that of Si) and the frequency of the double Dirac point is fairly low [35]. So, there is a possibility that we can employ an effective medium theory to describe the physics of the PC system. As the dispersions near the zone center are isotropic, the PC system can be described by three independent effective elastic parameters: the effective mass density $\rho_{eff}$, bulk modulus $\kappa_{eff}$, and shear modulus $\mu_{eff}$ which can be obtained using standard effective medium theory [36]. The results are plotted in Fig. 4, in which the red dashed-dotted, black dashed, blue solid and cyan dotted lines represent $\rho_{eff}$, $\kappa_{eff}$, $1/\mu_{eff}$ and $1/(\mu_{eff}+\kappa_{eff})$ as a function of frequency, respectively. Fig. 4 clearly shows that $\rho_{eff}$, $1/\mu_{eff}$ and $1/(\mu_{eff}+\kappa_{eff})$ intersects at zero at the double Dirac point ($\omega = 0.209(2\pi c_{t,Si}/a)$). For the eigenmode is a combination of quadrupolar and dipolar states only [35], $\kappa_{eff}$ does not exhibit resonant behaviors in the frequency region considered [36]. As $\rho_{eff}$, $1/\mu_{eff}$ and $1/(\mu_{eff}+\kappa_{eff})$ go through zero simultaneously and linearly, the effective refractive index for both the P wave and S wave also goes through zero but the group velocity remains finite.

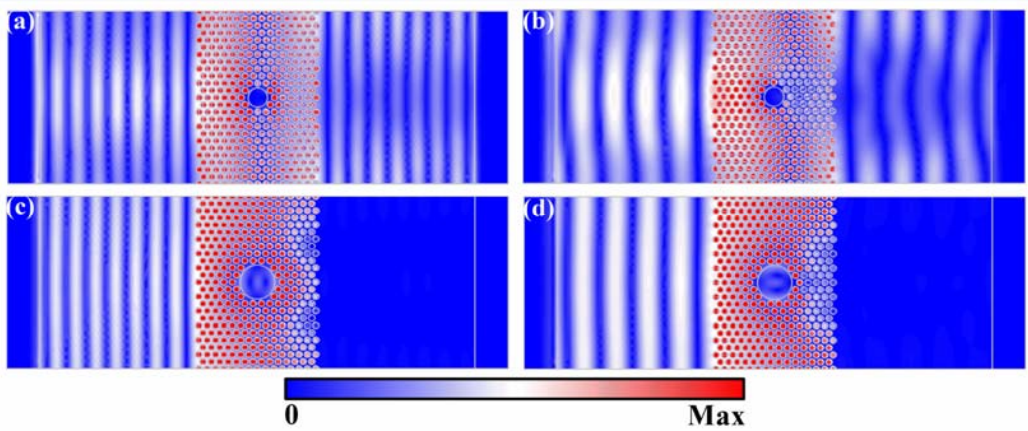

FIG. 5. Different columns represent the incident S and P waves, respectively. The numerically simulated displacement field distributions for the 2D PC system embedded with the steel cylinders to achieve transmission preserving plane-wave characteristic (panels (a) and (b)), and total reflection (panels (c) and (d)). The simulation domain is terminated in the propagation direction with PML.

Fig. 5 shows the results of numerical simulations which demonstrate the unusual wave propagation properties of the PC system. In Figs. 5(a) and 5(b), respectively, the displacement field distributions show that the incident S and P waves are able to pass through the PC system and still preserve their plane-wave characteristic when the embedded steel cylinder has radius of $r_d = 1.1a$. While in Figs. 5(c) and 5(d), the displacement field distributions of the PC system show the total reflection for the incident S and P waves when the steel cylinder has radius of $r_d = 2.06a$, satisfying the condition $E_{11}E_{22} - E_{12}E_{21} = 0$. Compared to previous scheme to manipulate elastic waves [15-18, 31], our PC system can achieve manipulation for in-plane S wave and P wave independently and simultaneously.

In conclusion, we show that, in the DZIM, there is no occurring of mode conversion for P wave or S wave incident on embedded objects, no matter whether there is occurring of wave scattering by the objects. A DZIM slab can be used either as a cloak of elastic waves for the embedded object when it is off-resonance, or as a blocker of elastic waves when the embedded object is on-resonance. A simple analytic model is presented to exhibit the essence of the physics uniformly. A 2D PC is suggested to achieve the intriguing phenomena. Our results provide new understanding of scattering of elastic waves by elastic discontinuities and enable novel way of controlling the propagation of elastic waves.

**Acknowledgments**

This work was supported by NSFC (Grant No.11304090 and 11174225) and the 973 Research Program of China (Grant No. 2015CB755500).

**References:**

[1] R. W. Ziolkowski, Phys. Rev. E **70**, 046608 (2004).

[2] S. Enoch, G. Tayeb, P. Sabouroux, N. Guerin, and P. Vincent, Phys. Rev. Lett. **89**, 213902 (2002).

[3] A. Alu, M. Silveirinha, A. Salandrino, and N. Engheta, Phys. Rev. B **75**, 155410 (2007).

[4] M. G. Silveirinha and N. Engheta, Phys. Rev. Lett. **97**, 157403 (2006).

[5] B. Edwards, A. Alù, M. E. Young, M. Silveirinha, and N. Engheta, Phys. Rev. Lett. **100**, 033903 (2008).

[6] J. Hao, W. Yan, and M. Qiu, Appl. Phys. Lett. **96**, 101109 (2010).

[7] V. C. Nguyen, L. Chen, and K. Halterman, Phys. Rev. Lett. **105**, 233908 (2010).

[8] Z. Liu, X. Zhang, Y. Mao, Y. Y. Zhu, Z. Yang, C. T. Chan, and P. Sheng, Science **289**, 1734 (2000).

[9] N. Fang, D. Xi, J. Xu, M. Ambati, W. Srituravanich, C. Sun, and X. Zhang, Nature Mater. **5**, 452 (2006).

[10] Y. Lai, Y. Wu, P. Sheng, and Z. Q. Zhang, Nat. Mater. **10**, 620 (2011).

[11] Y. Wu, Y. Lai, and Z. Q. Zhang, Phys. Rev. Lett. **107**, 105506 (2011).

[12] J. Christensen and F. Javier García de Abajo, Phys. Rev. Lett. **108**, 124301 (2012).

[13] A. Sukhovich, B. Merheb, K. Muralidharan, J. O. Vasseur, Y. Pennec, P. A. Deymier, and J. H. Page, Phys. Rev. Lett. **102**, 154301 (2009).

[14] C. M. Park, J. J. Park, S. H. Lee, Y. M. Seo, C. K. Kim, and S. H. Lee, Phys. Rev. Lett. **107**, 194301 (2011).

[15] G. W. Milton, M. Briane, and J. R. Willis, New J. Phys. **8**, 248 (2006).

[16] M. Brun, S. Guenneau, and A. B. Movchan, Appl. Phys. Lett. **94**, 061903 (2009).

[17] M. Farhat, S. Guenneau, and S. Enoch, Phys. Rev. Lett. **103**, 024301 (2009).

[18] N. Stenger, M. Wilhelm, and M. Wegener, Phys. Rev. Lett. **108**, 014301 (2012)

[19] S. A. Cummer and D. Schurig, New J. Phys. **9**, 45 (2007).

[20] A. N. Norris, Proc. R. Soc. A **464**, 2411 (2008).

[21] S. Zhang, C. Xia and N. Fang, Phys. Rev. Lett. **106**, 024301 (2011).

[22] F. Bongard, H. Lissek, and J. R. Mosig, Phys. Rev. B **82**, 094306 (2010).

[23] J. J. Park, K. J. B. Lee, O. B. Wright, M. K. Jung, and S. H. Lee, Phys. Rev. Lett. **110**, 244302 (2013).

[24] R. Fleury and A. Alu, Phys. Rev. Lett. **111**, 055501 (2013).

[25] Z. Liang and J. Li, Phys. Rev. Lett. **108**, 114301 (2012).

[26] Y. B. Xie, Bogdan-Ioan Popa, L. Zigoneanu, and S. A. Cummer, Phys. Rev. Lett. **110**, 175501 (2013).

[27] F. M. Liu, X. Q. Huang, and C. T. Chan, Appl. Phys. Lett. **100**, 071911 (2012).

[28] J. Mei, Y. Wu, C. T. Chan, and Z. Q. Zhang, Phys. Rev. B **86**, 035141 (2012).

[29] D. Torrent and J. Sanchez-Dehesa, Phys. Rev. Lett. **108**, 174301 (2012).

[30] Z. G. Chen, X. Ni, Y. Wu, C. He, X. C. Sun, L. Y. Zheng, M. H. Lu, and Y. F. Chen, Sci. Rep. **4**, 4613 (2014).

[31] F. M. Liu, Y. Lai, X. Q. Huang, and C. T. Chan, Phys. Rev. B **84**, 224113 (2011).

[32] T. Antonakakis, R. V. Craster, and S. Guenneau, Eur. Phys. Lett. **105**, 54004 (2014).

[33] K. F. Graff, *Wave Motion in Elastic Solids* (Dover Publications, New York, 1991).

[34] Z. Liu, C. T. Chan, and P. Sheng, Phys. Rev. B **71**, 014103(2005).

[35] See supplementary material for details of the independent control of P and S waves by using rectangular objects, the band structure and the eigenstates the proposed PC.

[36] Y. Wu, Y. Lai, and Z. Q. Zhang, Phys. Rev. B, **76**, 205313 (2007).